\documentstyle[twoside,epsfig]{adacta} 
\def\be{\begin{eqnarray}}
\def\ee{\end{eqnarray}}
\def\l{\langle}
\def\r{\rangle}

\begin{document} 

\prvastrana=177
\poslednastrana=184
\setcounter{page}{\prvastrana}

\def\autor{V. Bu\v{z}ek et al.}
\def\nazov{Flow of information in quantum cloners \dots}

\headings{177}{184}

\title{\uppercase{
Controlling the flow of  information in quantum cloners:
Asymmetric cloning 
}}
\author{V. Bu\v{z}ek$^{a c}$\footnote{\email{buzek@savba.sk}}, M.
Hillery$^{b}$, R. Bednik$^{c}$}
{${(a)}$ Institute of Physics, SLovak Academy of Sciences, D\'{u}bravsk\'{a}
cesta 9,\\ 842 28 Bratislava, Slovakia\\
${(b)}$ Department of Physics and Astronomy,\\ 
Hunter College of the City University of New York, \\
695 Park Avenue, New York, NY 10021, U.S.A.\\
$(c)$ Faculty of Mathematics and Physics, Comenius University, Mlynsk\'{a}
dolina F-1,\\ 842 15 Bratislava, Slovakia}

\datumy{15 May 1998}{29  May 1998}

\abstract
{We show that the distribution of information at the output of the quantum
cloner can be efficiently controlled via preparation of the 
quantum cloner. We present a universal cloning network with the help of
which asymmetric cloning can be performed.
}
\kapitola{1. Introduction}

An {\it unknown} pure state of a qubit can be ``swapped'' between
two parties (Alice and Bob) by a unitary transformation. To be specific,
let us assume that Alice has a qubit initially prepared in a 
pure quantum state $|\Psi\rangle_{a_0}$ 
\begin{eqnarray}
| \Psi\rangle_{a_0} = 
\alpha_0 |0\r_{a_0} + \alpha_1 |1\r_{a_0}
\label{1.1}
\end{eqnarray}
described as a vector in 
an $2$-dimensional Hilbert space ${\cal H}_{a_0}$
spanned by two orthonormal
basis vectors $|0\r_{a_0}$ and $|1\r_{a_0}$ . The complex
amplitudes $\alpha_i$ are normalized to unity, i.e.
$|\alpha_0|^2 + |\alpha_1|^2 =1$.
Simultaneously 
Bob has a qubit initially prepared in a specific
(i.e., known) state $|0\rangle_{a_1}$ which is a vector in the
 Hilbert space ${\cal H}_{a_1}$.
From the general rules of quantum mechanics it follows that there always
exists a {\it unitary} transformation $\hat{U}$ acting on 
${\cal H}_{a_0}\otimes
{\cal H}_{a_1}$ which swaps Alice's and Bob's states, i.e. 
\begin{eqnarray}
| \Psi\rangle_{a_0} |0\rangle_{a_1}\stackrel{\hat{U}}{\longrightarrow} 
|0\rangle_{a_0}| \Psi\rangle_{a_1}.
\label{1.2}
\end{eqnarray}
Operational meaning of this state swapping is as follows: \newline
{\bf (1)}  Prior the swapping
Alice does not know what is the state of her qubit. But, in
principle she can perform an optimal measurement on her system 
(see \cite{Helstrom76,Holevo82}) which would allow her 
to estimate the state. The quality of this estimation is characterized
by the mean fidelity ${\cal F}$ \cite{Massar95,Derka98}.
Taking into account that Alice has only
a single qubit, then the maximal value of the mean fidelity of the
estimation is ${\cal F}={2}/{3}$.
After the measurement is performed Alice can communicate classically
her result to Bob
(and, in fact, to an arbitrary number of recipients) who can recreate
the estimated state.
If this type of classical communication is not allowed by the rules of
the game, then  Bob can only guess what is the state (this estimation
via ``wild guessing'' corresponds to the minimal value of the mean
 fidelity, which in the case of a 
single qubit is ${\cal F}={1}/{2}$)
 \cite{Massar95,Derka98}.
Obviously, as soon as Alice performs the measurement 
the state $|\Psi\rangle_{a_0}$ is
``lost'', so there is nothing relevant to swap.
\newline
{\bf (2)} If Alice does not perform a measurement on her quantum system
she can quantum-mechanically 
swap the state $|\Psi\rangle$ to Bob. After the swapping she  
cannot gain information about $|\Psi\rangle$ (except, if Bob 
classically communicates any results of his measurements to her).

In this swapping scenario, when no classical communication is allowed,
{\it either} Alice {\it or}
 Bob has the state $|\Psi\rangle$ and it has to be decided {\it a priori}
(i.e., before the measurement) who is going to have the qubit (Alice or
Bob). At this point one can 
ask a question whether it would be possible to find a unitary transformation
such that both Alice and Bob would 
have the state $|\Psi\rangle$ {\it
simultaneously}. That is, the question is whether a unitary transformation 
$\hat{U}$ such that 
\begin{eqnarray}
| \Psi\rangle_{a_0} |0\rangle_{a_1}\stackrel{\hat{U}}{\longrightarrow} 
|\Psi\rangle_{a_0}| \Psi\rangle_{a_1},
\label{1.4}
\end{eqnarray}
does exist for an arbitrary (unknown) input state $|\Psi\rangle$.

Generalizing the proof of the Wootters-Zurek 
no-cloning theorem \cite{Wootters82}
it is easy to show that the linearity of  quantum mechanics
prohibits the existence of the perfect cloning expressed by Eq.(\ref{1.4}).
This is a major difference between quantum and classical information:
it is possible to make perfect copies of classical information,
but quantum information cannot be copied perfectly, i.e., quantum
states cannot be cloned perfectly. Nevertheless, if 
the requirement that the copies are perfect is dropped, then
it is possible to make quantum copies.  This was first shown in
Ref.~\cite{Buzek96}, where a transformation which produces two
mutually identical copies of an arbitrary input  
 qubit state  was given.  This transformation
was shown to be optimal, in the sense that it maximizes the average
fidelity between the input and output qubits, by Gisin and Massar
\cite{Gisin97} and by Bruss, et.\ al.\ \cite{Bruss98}.  
Gisin and Huttner \cite{Gisin97a} have shown that the quantum cloning can
be efficiently used for eavesdropping.  
Gisin and Massar
have also been able to find copying transformations 
which produce
$k$ copies from $l$ originals (where $k>l$) \cite{Gisin97}.
In 
addition, quantum logic networks for quantum copying machines 
of qubits have
been developed \cite {Buzek97b,Buzek97c,Buzek98a}, 
and bounds have been placed on
how good copies can be \cite{Hillery97,Bruss97}.
It has been shown recently \cite{Buzek97a} that the 
 inseparability of quantum states can be partially cloned (broadcasted)
with the help of local quantum cloning machines, i.e. distant parties
sharing an entangled pair of qubits can generate two pairs
of partially {\em nonlocally} entangled states using only {\em local}
operations. Gisin has presented an interesting proof \cite{Gisin98}
 of the optimality of the quantum cloner showing that the bound on
the fidelity of the universal quantum cloner 
 \cite{Buzek96}
is compatible with the no-signaling constraint.
Cerf \cite{Cerf98a,Cerf98b} 
has introduced a family of quantum cloning machines
that produce two approximate copies from a single qubit, while the
overall input-to-output operation for each copy is a Pauli channel.
Cerf has also introduced a concept of asymmetric quantum cloning
when at the output of the cloner the two clones are not identical, but
simultaneously they are specifically related to the original qubit (see
below). 
It has been shown in Ref.~\cite{Buzek98b} that states of quantum systems
in arbitrary-dimensional Hilbert spaces can be universally cloned
(i.e., the fidelity of cloning does not depend on the input). The
cloning transformation presented in \cite{Buzek98b} allows one to study 
how quantum registers can be cloned. It has been shown later by Werner
\cite{Werner98} that this cloning transformation is optimal. Moreover,
Werner in his elegant paper have constructed a universal transformation
for an optimal cloning 
which produces
$k$ copies from $l$ originals (where $k>l$) of an $M$ dimensional system.
Zanardi  \cite{Zanardi98} has presented a group-theoretical analysis
of the universal quantum cloning.

\podkapitola{1.1 The problem}

Let us assume the initial qubit to be in an {\it unknown} state
$\hat{\rho}_{a_0}$.
Our task is to clone this qubit universally, i.e. input-state independently,
in such a way, that we can control the scaling of the original and the
clone at the output. That is, 
we are looking for a cloner (the asymmetric cloner \cite{Cerf98a,Cerf98b})
in which we can control a flow of quantum
information in such a way that 
the two clones at the output can be represented as
\be
\hat{\rho}_{a_j}^{(out)} = s_j \hat{\rho}_{a_j}^{(id)} +\frac{1-s_j}{2}\hat{1},
\label{1}
\ee
where $j=0,1$. Here we assume that the original qubit after the cloning
is ``scaled'' by the factor $s_0$, while the copy is scaled by the
factor $s_1$. These two scaling parameters are not independent and they
are related by a specific inequality (see below). We note the two extreme
cases, when {\bf (a)} $s_0=1$ and $s_1=0$ and, vice versa, when {\bf (b)}
$s_0=0$ and $s_1=1$. These correspond to the following situations: {\bf (a)}
 the information 
is completely preserved in the original qubit, and  {\bf (b)}
the information is totally transfered (swapped) to the copy. 
Symmetric cloning corresponds to the situation when $s_0=s_1$.  

Our main task in this
paper is to find a  cloning network in which the control over the
flow of information (i.e. the control over the values of the scaling
parameters $s_0$ and $s_1$) can be performed via 
 preparation of the initial state of the  cloner.

\kapitola{2. Network for asymmetric cloner}

For simplicity, let us assume that the original qubit is initially
in a pure state (\ref{1.1}),
 i.e. $|\Psi\rangle_{a_0}^{(in)}=\alpha_0|0\r+\alpha_1|1\r$.
To perform asymmetric cloning we have to unitarily couple
to the original qubit with two additional qubits denoted as $a_1$ and $b_1$
which are initially  in a pure state $|0\r_{a_0}\otimes|0\r_{b_1}
\equiv |00\r$. 

At the first stage of the  cloning
these two qubits are transformed from the state $|00\r$ into the state
\be
|\Psi\r_{a_1b_1}^{(prep)}=C_1|00\r + C_2 |01\r + C_3 |10\r + C_4|11\r,
\label{2}
\ee
where the complex amplitudes $C_i$ will be specified so that the conditions
given by Eq.(\ref{1}) are fulfilled\footnote{We do not specify
this preparation part of the cloning network (see Fig.~1) because
it is well know that the state (\ref{2}) can be prepared from $|00\r$
via a simple sequence of local operations and C-NOT gates \cite{Barenco}.}.
 At this stage the original qubit is
still not involved in the process of cloning, but the choice of $C_i$'s
later affects the flow of information in the cloner.

\begin{figure}[t]
\leavevmode
\epsfxsize=12cm 
\centerline{\epsfbox{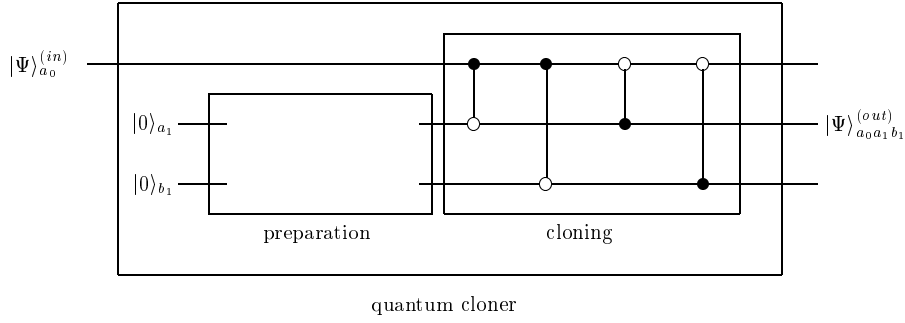}}

\medskip

\caption{Fig.~1.
Graphical representation of the  
 network for the asymmetric cloner.
  The logical
controlled-NOT $\hat{P}_{kl}$ given by Eq.(\ref{3})
has as its input a control qubit
(denoted as $\bullet$ ) and  a target qubit  (denoted as $\circ$ ).
We separate the preparation of the quantum copier from the
cloning process itself. The cloning, i.e. the transfer of quantum
information from the original qubit, is performed by a sequence of
four controlled-NOTs. 
We do not specify the details of the preparation part of the network
which generates the state $|\Psi\r^{(prep)}_{a_1b_1}$ out of the
input $|00\r_{a_1b_1}$. This is a well known network for generation
of an arbitrary two-qubit state.
}
\label{fig1}
\end{figure}

After the preparation stage 
we assume that the
interaction between the original
qubit and two additional qubits $a_1$ and $b_1$ is performed via a
simple sequence of four C-NOT gates (see Fig.~1).
The operator which implements the C-NOT gate, $\hat{P}_{kl}$,
acts on the basis vectors of the two qubits as follows ($k$
denotes the control qubit and $l$ the target):
\begin{eqnarray}
\begin{array}{rcl}
\hat{P}_{kl} | 0 \rangle_k | 0 \rangle_l & = &
| 0 \rangle_k | 0 \rangle_l;\\
\hat{P}_{kl} | 0 \rangle_k | 1 \rangle_l & = &
| 0 \rangle_k | 1 \rangle_l;\\
\hat{P}_{kl} | 1 \rangle_k | 0 \rangle_l & = &
| 1 \rangle_k | 1 \rangle_l;\\
\hat{P}_{kl} | 1 \rangle_k | 1 \rangle_l & = &
| 1 \rangle_k | 0 \rangle_l.
\end{array}
\label{3}
\end{eqnarray}

We assume the specific action of  four controlled-NOT operations 
\begin{eqnarray}
|\Psi\rangle_{a_0a_1b_1}^{(out)} = \hat{P}_{b_1a_0} \hat{P}_{a_1a_0}
\hat{P}_{a_0b_1}
\hat{P}_{a_0a_1}|\Psi\rangle_{a_0}^{(in)}  |\Psi\rangle_{a_1b_1}^{(prep)},
\label{4}
\end{eqnarray}
(see Fig.~1)
during
which the information is transfered from $a_0$ qubit to other
two qubits. 
From the state vector $|\Psi\rangle_{a_0a_1b_1}^{(out)}$ 
given by Eq.(\ref{4}) 
we obtain  single-qubit density operators
\be
\hat{\rho}_{a_0}^{(out)} & =  {\rm Tr}_{a_1b_1}\left[ 
|\Psi\rangle_{a_0a_1b_1}^{(out)}\l\Psi|\right]; 
\nonumber
\\
\hat{\rho}_{a_1}^{(out)} & ={\rm Tr}_{a_0b_1}\left[ 
|\Psi\rangle_{a_0a_1b_1}^{(out)}\l\Psi|\right],
\label{5}
\ee
which explicitly depend on the complex amplitudes 
$C_j=c_j {\rm e}^{i\theta_j}$ (here $c_j=|C_j|$). These amplitudes 
come into play via the preparation of the state 
$|\Psi\rangle_{a_1b_1}^{(prep)}$ (\ref{2}).
Our task now is to specify these four  amplitudes 
so that the density operators (\ref{5}) fulfill the scaling 
condition (\ref{1}). Comparing Eqs.(\ref{1})
and (\ref{5}) we find that the density operators (\ref{5}) can be
written in the scaled form (\ref{1}) if the complex amplitudes
$C_j$ and the two scaling factors $s_0$ and $s_1$ are related as
\be
c_1=\sqrt{\frac{s_0+s_1}{2}};\qquad
c_2=\sqrt{\frac{1-s_0}{2}};\qquad
c_4=\sqrt{\frac{1-s_1}{2}},
\label{6}
\ee
and 
\be
\cos(\theta_1-\theta_2)=
\frac{s_1}{\sqrt{(s_0+s_1)(1-s_0)}};
\nonumber
\\
\cos(\theta_1-\theta_4)=
\frac{s_0}{\sqrt{(s_0+s_1)(1-s_1)}};
\label{7}
\ee
while $C_3=0$. With these complex amplitudes $C_j$ the quantum network
as described by Eq.(\ref{4}) realizes the asymmetric cloner. From
Eqs.(\ref{6}) and (\ref{7}) we find that the scaling factors $s_0$ and 
$s_1$ have to be related as (see Fig.~2)
\be
s_0^2+s_1^2+s_0s_1-s_0-s_1\leq 0.
\label{8}
\ee

\begin{figure}[t]
\leavevmode
\epsfxsize=5cm 
\centerline{\epsfbox{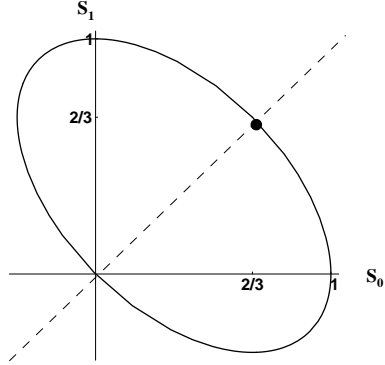}}

\medskip

\caption{Fig.~2.
The ellipse delimiting the range of possible value of the scaling
parameters $s_0$ and $s_1$ of the two clones that simultaneously
emerge as outputs
of the asymmetric cloner. We see that the symmetric
universal cloner corresponds to $s_0=s_1=2/3$.
}
\label{fig2}
\end{figure}

To understand more clearly how the asymmetric cloner works we assume
three specific preparation states (\ref{2}) of the cloner:\newline
{\bf (i)} Let us assume that the cloner is initially prepared in the
maximally entangled two-qubit (Bell) state  
\be
|\Psi\r_{a_1b_1}^{(prep)}=\frac{1}{\sqrt{2}}(|00\r + |11\r),
\label{8a}
\ee
which can be prepared from $|00\r$ 
 via a simple sequence of the Hadamard transformation
on the qubit $a_1$ followed by the C-NOT operation with $a_1$ being the
control. It easy to see that the cloner which is prepared in the state
(\ref{8a}) does not affect  the original qubit 
which at the output is in the same state as in the input, i.e.
 we find that $s_0=1$ and $s_1=0$.
\newline
{\bf (ii)} The cloner, which is initially in the completely disentangled state
\be
|\Psi\r_{a_1b_1}^{(prep)}=|0\r_{a_1}\otimes
\frac{1}{\sqrt{2}}(|0\r + |1\r)_{b_1},
\label{8b}
\ee
acts as the state swapper, i.e. in this case $s_0=0$ and $s_1=1$,
which means that the initial state of the original qubit $a_0$ 
is ``unitarily teleported'' to the qubit $a_1$.
The state (\ref{8b}) can be obtained from $|00\r$
by the action of the Hadamard transformation on the qubit $b_1$.
\newline
{\bf (iii)} The optimal universal quantum cloning \cite{Buzek96}
of the original qubit
(i.e. $s_0=s_1=2/3$)
can be realized when the cloner is initially prepared in the state  
\be
|\Psi\r_{a_1b_1}^{(prep)}=
\sqrt{\frac{2}{3}}|00\r + \frac{1}{\sqrt{6}}(|01\r + |11\r). 
\label{8c}
\ee
This state can be prepared with the help of a simple network presented
in Ref.\cite{Buzek97c}.

\kapitola{3. Instead of conclusions: Pauli cloners}

We have presented a simple logical network with the help of
which asymmetric cloning of qubits can be performed. This 
network is suitable for 
cloning of pure as well as impure  input states. In fact,
an impure state of a qubit can be represented as
a state of a  subsystem of a composite
system of two qubits. This composite two-qubit 
system itself is assumed to be in a pure state. Let us therefore consider
cloning of the initial qubit $a_0$ which is initially entangled with
a reference qubit $r$. To be specific, let us assume that the two qubits
 are prepared initially in the maximally entangled Bell state
$|\Phi^+\r_{ra_0}$.
Here the four maximally entangled states of two qubits are defined as
usually
\be
|\Phi^{\pm}\rm\r=\frac{1}{\sqrt{2}}(|00\r\pm|11\r);
\nonumber
\\
|\Psi^{\pm}\rm\r=\frac{1}{\sqrt{2}}(|01\r\pm|10\r).
\label{9}
\ee
Let us assume that the cloner is initially prepared in the state 
\be
|\Psi\rangle_{a_1b_1}^{(prep)}=
X_1|\Phi^+\r +X_2|\Phi^-\r + X_3|\Psi^+\r + X_4|\Psi^-\r,
\label{10}
\ee
and we apply the sequence 
\begin{eqnarray}
|\Psi\rangle_{ra_0a_1b_1}^{(out)} = \hat{P}_{b_1a_0} \hat{P}_{a_1a_0}
\hat{P}_{a_0b_1}
\hat{P}_{a_0a_1}|\Phi^+\rangle_{ra_0}^{(in)}
|\Psi\rangle_{a_1b_1}^{(prep)},
\label{11}
\end{eqnarray}
of four C-NOT's on the qubits
$a_0$ and $a_1b_1$ (see 
Fig.~3).
\begin{figure}[t]
\leavevmode
\epsfxsize=12cm 
\centerline{\epsfbox{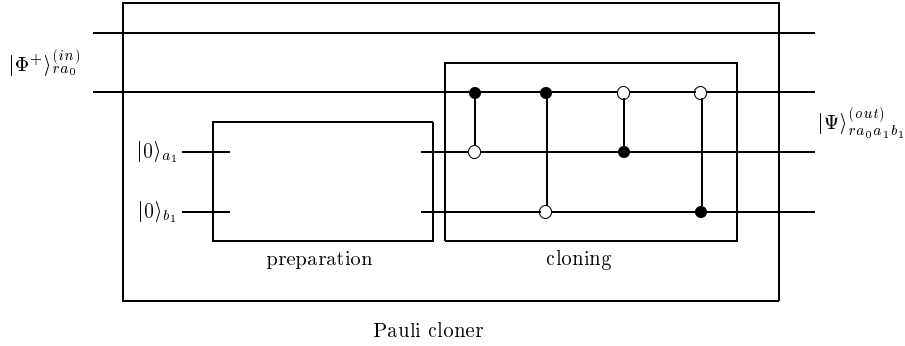}}

\medskip

\caption{Fig.~3.
Graphical representation of the  
 network for the Pauli cloner.
  The logical
controlled-NOT $\hat{P}_{kl}$
are specified as in Fig.~1.
Here again, the cloning, i.e. the transfer of quantum
information from the original qubit, is performed by a sequence of
four controlled-NOTs. 
}
\label{fig3}
\end{figure}
The 4-qubit state $|\Psi\rangle_{ra_0a_1b_1}^{(out)}$ at the output
reads
\be
|\Psi\rangle_{ra_0a_1b_1}^{(out)}=
\left\{ X_1|\Phi^+\r|\Phi^+\r 
+X_2|\Phi^-\r|\Phi^-\r
+X_3|\Psi^+\r|\Psi^+\r
+X_4|\Psi^-\r|\Psi^-\r\right\}_{ra_0;a_1b_1},
\label{12}
\ee
which means that the network (\ref{11})  realizes the 
Pauli cloner introduced recently by Cerf \cite{Cerf98b}.

\medskip
\noindent {\bf Acknowledgements}
One of us (V.B.) thanks Nicolas Cerf for helpful discussions and
correspondence.
This research was supported by a grant from the Royal Society.

\medskip
\noindent{\sl Note added} After this paper was completed
a paper by Niu and Griffiths \cite{Niu98} appeared in the
Los Alamos e-print archive in which asymmetric cloning is studied
from a different perspective 
and alternative cloning networks are presented.


\end{document}